**Title** Polarization-dependent conductivity of grain boundaries in BiFeO3 thin films

*Author(s), and Corresponding Author(s)\**
*Denis Alikin\*, Yevhen Fomichov, Saulo Portes Reis, Alexander Abramov, Dmitry Chezganov, Vladimir Shur, Eugene Eliseev, Anna Morozovska\*, Eudes Araujo, and Andrei Kholkin\**

Dr. D.O. Alikin[1,2], Dr. Y. Fomichov[3], Dr. S.P. Reis[4,7], A.S. Abramov[1], Dr. D.S. Chezganov[1], Prof. V.Ya. Shur[1], Dr. E. Eliseev[5], Dr. A. Morozovska[6], Dr. E.B. Araujo[4], and Dr. A.L. Kholkin[1,2]

[1]School of Natural Sciences and Mathematics, Ural Federal University, 620100, Ekaterinburg, Russia
[2]Department of Physics & CICECO-Aveiro Institute of Materials, 3810-193, University of Aveiro, Portugal
[3]Faculty of Mathematics and Physics, Charles University in Prague, Prague 8, 180 00, Czech Republic
[4]Department of Chemistry and Physics, São Paulo State University, Ilha Solteira - SP, Brazil
[5]Institute for Problems of Materials Science, National Academy of Sciences of Ukraine, 03142 Kyiv, Ukraine
[6]Institute of Physics, National Academy of Sciences of Ukraine, 03028 Kyiv, Ukraine
[7]Federal Institute of Education, Science and Technology of São Paulo, 15503-110 Votuporanga, Brazil

E-mail: kholkin@ua.pt, denis.alikin@urfu.ru, anna.n.morozovska@gmail.com

**Abstract**

Charge transport across the interfaces in complex oxides attracts a lot of attention because it allows creating novel functionalities useful for device applications. In particular, it has been observed that movable domain walls in epitaxial BiFeO$_3$ films possess enhanced conductivity that can be used for read out in ferroelectric-based memories. In this work, the relation between the polarization and conductivity in sol-gel BiFeO$_3$ films with special emphasis on grain boundaries as natural interfaces in polycrystalline ferroelectrics is investigated. The grains exhibit self-organized domain structure in these films, so that the "domain clusters" consisting of several grains with aligned polarization directions are formed. Surprisingly, grain boundaries between these clusters (with antiparallel polarization direction) have significantly higher electrical conductivity in comparison to "inter-cluster" grain boundaries, in which the conductivity was even smaller than in the bulk. As such, polarization-dependent conductivity of the grain boundaries was observed for the first time in ferroelectric thin films. The results are rationalized by thermodynamic modelling combined with finite element simulations of the charge and stress accumulation at the grain boundaries giving major contribution to conductivity. The observed polarization-dependent conductivity of grain boundaries in ferroelectrics opens up a new avenue for exploiting these materials in electronic devices.

## 1. Introduction

Many efforts have been devoted so far to achieve the control of interfaces in ferroelectric materials based on their polarization. These efforts resulted in the discovery of a variety of different phenomena such as polarization-dependent tunneling effect, resistive switching, symmetry breaking, etc. [1,2] In particular, domain wall conductivity[3–5], formation of topological defects[6,7], phase boundaries[8] and ferroelectric-insulator interfaces[9] have been studied. In general, the control of the local conductivity along these interfaces can be engineered based on the mutual orientation of adjacent polarization states [7,9]. However, to the best of our knowledge no studies of the polarization-dependent conductivity in more complex grain boundary (GB) interfaces in ferroelectric materials have been undertaken so far.

It is well known the macroscopic properties of polycrystalline ferroelectrics are significantly entangled by their structural heterogeneity caused by not only by the existence of domains and domain walls, but also by complicated grain and phase boundary interfaces, large macroscopic defects, such as dislocations etc. Due to this complexity the domain wall contribution to the functional properties such as dielectric constant and piezoelectric coefficient (so-called extrinsic contribution) was only approximately estimated and it was thought to be about 50-70 % of the total response.[10,11] GBs are the interfaces between crystallites with different crystallographic orientations and have been of the main interest because they are responsible for the pinning of domain walls that naturally decreases their mobility. [12,13] GBs are difficult to control because they are formed at the stage of materials synthesis at elevated temperature, when the accelerated diffusion of mobile species occurs during grain growth and their mechanical consolidation. At room temperature they become stable and represent mostly 2D defects caused by the structural and chemical disorder.[14,15] GBs' electronic and ionic transport properties are different from the bulk due to large concentration of structural defects and impurities.[14,16] Thereby, they impact directly on the leakage currents, breakdown strength, dielectric permittivity and piezocoefficients of the ferroelectric polycrystalline materials (e.g., may act as conductive inclusions effectively influencing dielectric permittivity or potential barriers for electronic or ionic transport).[17–19] On the other hand, it is well known that GBs are one of the key factor preventing domain wall motion in polycrystalline materials.[20–23] It has been shown that the increase in the apparent grain size could activate larger extrinsic contributions to the dielectric permittivity and piezoelectric effect[24] thus leading to apparent grain size dependence.

Another issue is the effect of conductivity and dielectric permittivity of GBs on the leakage current and dielectric response of the bulk. The segregation of the defects and/or impurities in vicinity of GBs appears at elevated temperature is known to provide a discontinuity of chemical potential and to counterbalance intergranular strain.[25–27] Generally accepted model is that GBs consist of "core" regions containing charged defects and thus exhibiting positive or negative charges. Nearby the core, the space charge regions of about 30-50 nm in thickness form that are accumulated\depleted by the majority charge carriers.[15,28] The existence of space charge strongly modifies the electronic transport across GBs, which can be described by the so-called "GB limited conduction" model.[14,29]. The conductivity macroscopically measured across and along GBs is dependent both on both bulk conductivity mechanism and chemistry of the defects segregated at the GBs.[15] Generally, the majority charge carriers of GB core have the same sign as the bulk ionic majority defects. For example, in ferroelectric $BaTiO_3$ (BTO), similar to many other ionic polycrystalline conductors, the conductivity along GBs is lower than that in the bulk for both acceptor and donor doped samples, while the sign of bulk majority defect conserves.[15] The electronic conduction mechanism along the GBs was suggested to be the hopping of localized electrons via charged defects.[30] In the popular perovskite material BiFeO$_3$ (BFO) the domain walls were found to be more conductive than the bulk.[3–5] This effect was rationalized by the local decrease of the bandgap under the action of free charge carriers trapped at the domain walls.[3–5] Later, strong angular dependence of the carrier accumulation was suggested to originate from the local band bending via angle-dependent electrostriction and flexoelectric coupling mechanisms.[31] Further, the enhanced conductivity was observed not only at the domain walls, but also at other interfaces such as grain and phase boundaries, secondary phases, polar-nonpolar interfaces etc. [12,19,32] Still an open question is a relation between the bulk conductivity of single- and polycrystalline materials and the local conductivity across the interfaces. Epitaxial films and single-crystals of BFO were shown to be *n*-type electronic conductors with the domain walls of the same conductivity type. However, the transport mechanism in these materials is still under discussion. [25,33] Recent microscopic studies revealed the existence of Bi vacancies segregation near domain walls of different types[25], which was suggested to enhance the hole conductivity through the interfaces. Contrary to this, Schrade et al. suggested the pure BFO to possess different mechanisms of electronic and ionic transport realized through the interior of the grains and across interfaces.[33] However, in a number of reports no significant conductivity along the GBs in polycrystalline BFO-based films has been observed. [34–36]

In this work, we study local polarization and conductivity distribution inside the grains and at the GBs in BFO films prepared by sol-gel technique in order to uncover the mechanism of their interrelation in polycrystalline ferroelectric materials. Though in widely accepted models of charge transport across GBs the polarization discontinuity at GBs is not taken into account, the variation of space charge concentration at the GBs is expected to depend on the spontaneous polarization in the adjacent grains.[28] We found that the polarization in these grains is self-organized in mesoscale clusters with uniform polarization comprising several grains. The GBs between these clusters have significantly higher electrical conductivity in comparison with the "inter-cluster GBs", in which conductivity was even smaller than in the bulk. These new effects are thought to be important for the design and application of ferroelectric materials in the thin film form.

## 2. Results and discussion

### *2.1. Dielectric relaxation*

So that to understand the effect of the GBs, we first studied the macroscopic dielectric response of our sol-gel BFO films. **Figure 1** presents the real part of the dielectric permittivity ($\varepsilon'$) and dielectric loss ($\tan \delta = \varepsilon''/\varepsilon'$) of BFO thin films as a function the frequency ($10^2 - 10^6$ Hz) at room temperature. The real part of permittivity is almost dispersion-free within the studied frequency range while a pronounced increase in the dielectric loss with decreasing frequency occurs below approximately 10 kHz. Despite slightly higher permittivity observed at 100 Hz ($\varepsilon' = 91$), the dielectric permittivity values $\varepsilon' = 78 \pm 2$ in the range 100 kHz-1 MHz are in a close agreement with the results reported earlier for randomly oriented BFO thin films.[28] Although $\tan \delta$ increases drastically at low frequencies, the dielectric loss above 100 kHz is relatively low (3.93% at 100 kHz and 2.5% at 1 MHz) and again matches the values reported in the literature.

The increase of the dielectric loss at low frequencies indicates an approach to dielectric relaxation with a peak in $\varepsilon''$ expected below 100 Hz. Such a behavior could be due to electron-hole hopping mechanism of the dielectric response.[33,37,38] Alternatively, the increase in the dielectric permittivity can arise from the contribution of conductive domain walls and interfaces via a Maxwell-Wagner mechanism.[19]

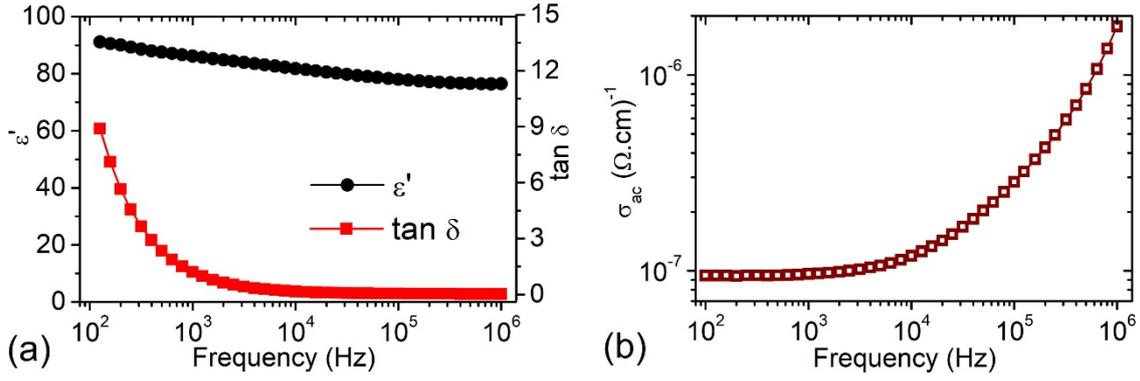

Figure 1. Frequency dependencies of real dielectric permittivity (ε´) and dielectric loss (tan δ) (a) and ac conductivity (b) of BiFeO$_3$ thin films at room temperature.

**Figure 2(a)** shows the frequency dependence of the real and imaginary parts of the complex impedance of the studied BFO films at room temperature while **Figure 2(b)** represents the variation of real and imaginary parts of the complex electric modulus as a function of frequency. Nyquist plots of the complex impedance and the complex electric modulus of BFO films are shown in **Figures 2(c)** and 2(d), respectively. In **Figure 2(b)**, an almost sigmoid behavior is observed in the real part of electric modulus $M'$ with its value tending to zero at around 100 Hz. This behavior indicates that the electrode polarization is negligible in the studied BFO films. *DC* conductivity ($\sigma_{dc}$), relaxation frequencies ($f_{max}$) and the $\alpha$ parameter obtained from theoretical fits are summarized in the Supporting Information B (Table B1). The obtained $\alpha \sim 0.98$ from the impedance (Z´ and Z´´) and the electric modulus (M´ and M´´) plots were almost the same indicating a small deviation from the ideal Debye-type relaxation.

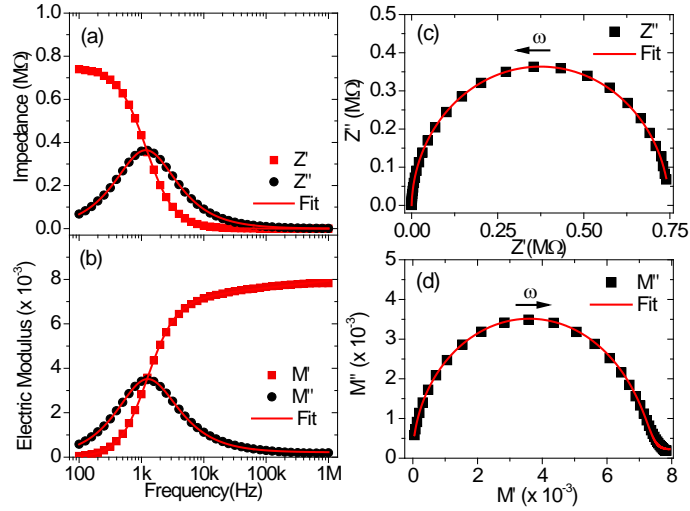

Figure 2. Frequency dependence of (a) real (Z´) and imaginary (Z´´) parts of impedance and (b) real (M´) and imaginary (M´´) parts of electric modulus of BiFeO$_3$ thin film. Nyquist plots of (c) complex impedance and (d) complex electric modulus of BiFeO$_3$ film made at room temperature. The continuous lines are theoretical approximation by equations of Cole-Cole model (see Supporting Information for details).

The frequency dependencies of the imaginary parts of impedance and electric modulus of the studied BiFeO$_3$ thin films were found to be close to each other with an appreciable mismatch, without a perfect overlap (see supporting information B). The overlapping peak positions of $Z''_{max}$ and $M''_{max}$ provides the evidence of delocalized or long-range relaxation while appreciable separation between these peaks indicates the presence of localized charge carriers. The observed slight difference in the peak position suggests the coexistence of contributions from both localized and long-range relaxations in the studied films. Following Rojac et al.[25] and Schrade et al.[33] we can interpret such a behavior as due to the electron-hole hopping between the sites with different valence state (Fe$^{3+}$-Fe$^{4+}$ or Bi vacancies). **Figure 1(b)** shows the *ac* conductivity as a function of frequency at room temperature for the studied films. In this Figure, the frequency independent plateau-like region observed at low frequencies is attributed to the dc conductivity $\sigma_{dc}$, which is expected to be frequency-independent at ω→0. The conductivity remains almost independent of frequency up to ~ 1 kHz and then increases with increasing frequency. This behavior is well known for ferroelectric semiconductors and again confirms the existence of hopping conductivity via lattice defects.[37]

The activation energies extracted from the temperature dependences of dielectric permittivity (data not shown here) are 0.42 eV and 0.43 eV for grain interior and GBs, respectively. These value match well the activation energy for *p*-type conductivity via hole transport reported in the past.[39] Usually Nyquist plots should reveal two distinct semicircle arcs of comparable sizes when the specific conductivity of the bulk is comparable with that of GBs. In our case, only one semicircle is observed, which is attributed to the domination contribution of the bulk conductivity[15]. In the present work, the bulk resistance effect is almost suppressed by the resistance of the GBs in **Figure 2c,d** because $\sigma_b \sim 10^3\ \sigma_{gb}$.

Nevertheless, it must be noted that the dielectric spectroscopy could be used to distinguish the resistive GBs (less conductive than the bulk), while GBs with enhanced conductivity cannot be well identified. This is because the conductive bulk and the GBs are connected in parallel and the interfacial capacitance is negligible, while the bulk capacitance is dominant. It is important to note, that *dc* conductivity of the BFO films reported here is almost ~ 30 times higher than in BFO thin films deposited by pulsed laser deposition.[28] As such, highly conductive GBs in BFO films were expected to have a significant effect on $\sigma_b$. To uncover the charge transport mechanisms, local conductivity measurements via high resolution conductive atomic force microscopy (c-AFM) were undertaken as described in the following section.

### 2.2. Polarization clusters and domain formation

In order to distinguish microstructural source of the dielectric relaxation, we studied domain structure and local conductivity by vector PFM and c-AFM, respectively. We observed clear piezoresponse images both in lateral and vertical PFM signals (**Figure 3**). That is expected for the multiaxial material with rhombohedral crystal symmetry. We found that the individual grains were all single domain and did not contain any domain walls inside the grains. It is well known that the domain size in the individual grains should depend on the grain size via so-called Kittel's law, which was established in a full analogy with magnetic systems.[40,41] As in the magnetics' case the critical size of the single domain grains is expected to exist where superparaelectric state is observed ('single domain –single grain').[42] This is in line with other materials prepared by sol-gel route: $Pb(Zr_xTi_{1-x})O_3$ and $Bi_{3.35}La_{0.85}Ti_3O_{12}$ films.[43–45] However, the critical size was larger than 10-20 nm, the value expected from the Kittel's theory applied to ferroelectrics[42]. Moreover, single domain state of individual grains could be broken by inhomogeneous electric field of biased atomic force microscopy (AFM) tip (Supporting Information C, Figure C1). This hints to the so-called "non-ergodic" superparaelectricity [42], in which the potential barrier for domain formation is lower and can be overcome by the action of high amplitude electric field of the AFM tip.

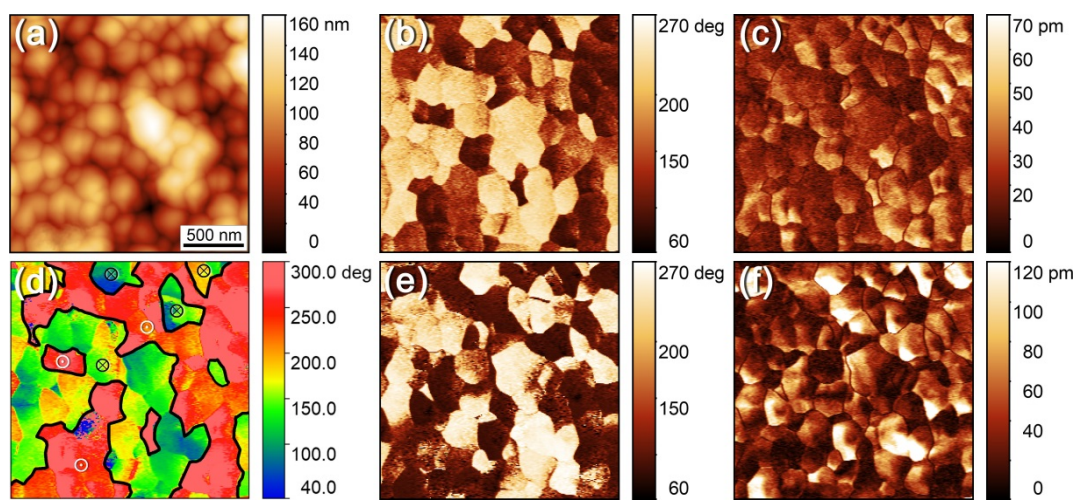

Figure 3. Ferroelectric domain structure in $BiFeO_3$ thin film: (a) topography, (b) vertical PFM phase, (c) vertical PFM amplitude, (d) vector PFM phase extracted according to Ref.[46] with boundaries of the clusters dedicated by black lines and out-of-plane polarization direction dedicated by corresponding symbols, (e) lateral PFM phase, (f) lateral PFM amplitude.

Surprisingly, we found self-assembled arrangement of the domain structure confined in "clusters" with the correlated orientation of the spontaneous polarization. This follows from the distribution of the vertical and lateral PFM phase signals that is not random. The reconstructed according to Ref. [46] vector PFM image is represented at **Figure 3d**, where clusters with the same sign of the piezoresponse (i.e. up or down polarization directions) are highlighted. Vector PFM approach combines both in-plane and out-of-plane PFM signal to recover dominant orientation of polarization vector in azimuthal plane[46]. From comparison of **Figures 3b, 3d and 3e**, it is seen that vector PFM phase contrast mostly repeats out-of-plane contrast that manifests dominant orientation of polarization in out-of-plane direction. Such preferably oriented clusters usually consisted of 3-20 grains. It must be noted that the cluster size varied across the sample, which is indicative that the driving force for unipolar state (e.g., internal bias field leading to self-polarization effect [47]) is non-uniform across the sample.

Analysis of electron backscattering diffraction patterns as well revealed non-uniform distribution of Euler angle at the poled figures with preferable orientation in the out-of-plane direction (supporting information D, figure D1, figure D2). At the same time no apparent mechanical texturing pattern can be found[48]. This demonstrates absence of preferable crystallographic direction for the individual grains. As such, observed self-organization can be referred to "electrical texturing" or, so-called, "self-poling" effect. Self-poling was earlier revealed in BFO thick films (tens of microns thick)[49]. It was proposed that the strain gradient appearing during cooling down from annealing temperature as a main mechanism of the domain arrangement. Other mechanisms were also mentioned such as flexoelectric poling and redistribution of charged defects[49]. It must be noted that formation of domain clusters can be called local 'self-poling' effect because preferred orientation of spontaneous polarization occurs only at the mesoscale.

### 2.3. Local conductivity "inter-cluster" and "intra-cluster" grain boundaries

Intuitively, it is expected that the GBs in the interior of the cluster and at its circumference would have different properties due to electrically different boundary conditions for the spontaneous polarization. This suggestion was verified via current measurements at the nanoscale (**Figure 4**). The GBs at the cluster circumference were highly conductive, while the electrical conductivity of GBs localized inside the clusters was equal or even lower than that in the bulk.

The phenomena of enhanced conductivity at the polar interfaces of different nature has been discovered in the contact between two insulating oxides[50,51], charged domain walls in ferroelectrics[52–54] and even at the virgin ferroelectric surfaces free from adsorbates and contamination[55]. The value of the enhanced current and properties of 2d electron gas were polarization dependent and polarization reversal could be used to create and erase the resulting conductive states[52,53]. We should, therefore, expect that the properties of such conductive GBs could be influenced by the corresponding domain state that eventually lead to the change of the conductivity distribution. Surprisingly, it was not the case. Even after complete uniform polarization switching by PFM tip the conductivity distribution did not change (see Supporting Information C, Figure C2). We assume that significant enhancement of the electronic conductivity at such GBs can be one of the source of high DC electrical conductivity of sol-gel films revealed by dielectric spectroscopy.

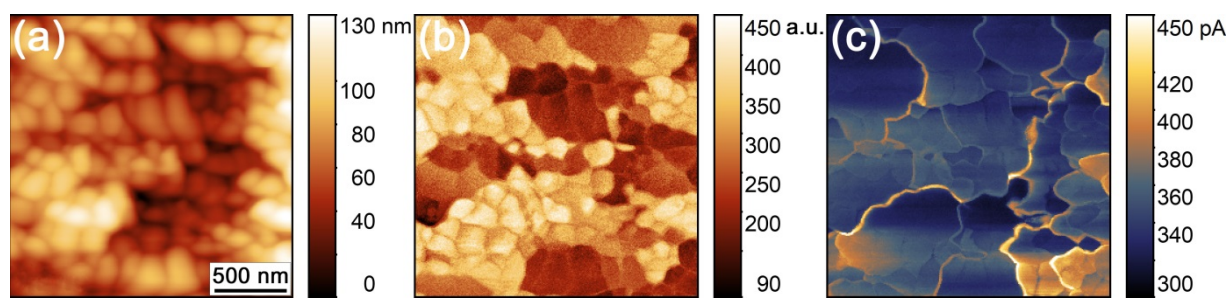

Figure 4. High and low conductive grain boundaries in BiFeO$_3$ thin film: (a) topography, (b) PFM response, (c) electric current. It must be noted that at Figure 4 we used current image done with additional photo-illumination in order to have better signal-to-noise ratio of the images allowing to distinguish details of the current distribution. The images obtained in dark conditions are provided in supporting information E.

Here we have to mention that the idea of two types of GBs was hypothesized a long time ago by Belyaev et al.[56] They studied lead titanate-zirconate, barium titanate, and bismuth titanate sintered in a temperature gradient. A significant difference of the dielectric permittivity, piezoelectric coefficient and conductivity (the latter has not been shown in Ref. [56]) being measured in the direction perpendicular and along the temperature gradient was found. The grains in these ceramics form cluster with different type of GBs inside and outside the cluster. Different types of GBs were suggested to have different misorientation and defectiveness. These results are in line with what we observed in our sol-gel BFO films (**Figure 5**). Often, clusters in BFO films demonstrate a visible separation in topography with pronounced cracks between different crystallites that we attribute to the effect of mechanical stress at the sintering stage (**Figure 3a**). Exactly these positions in the studied sample possessed enhanced conductivity (**Figure 3b,c**).

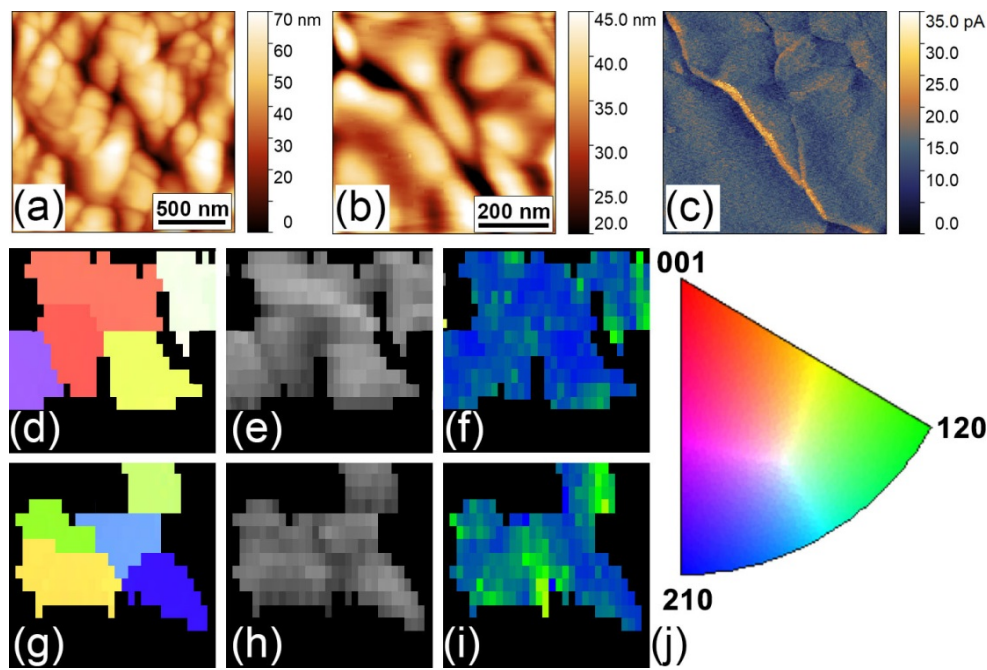

Figure 5. (a) and (b) Topography, (c) C-AFM and (d)-(i) EBSD images of grain agglomerates in polycrystalline BFO sol-gel thin films (5·5 μm size). (d)-(f) and (g)-(i) two different places across the surface. (d), (g) inverse polar figure (IPF) Z maps, (e), (h) image quality (band contrast) map, (f),(i) local misorientation map. (j) Legend for IPF Z contrasts at (d),(g). The relative intensities at image quality and local misorientation maps are chosen equally for reliable comparison.

In the locations where mechanical stress was maximal during cooling in gradient conditions the strain is expected to appear. Here we used EBSD to study the grain agglomerates with preferable orientation in out-of-plane direction GBs across the sample (**Figure 5d-i**). It must be noted that, due to imperfect indexing, only some places in the films could be resolved. Thereby, our further discussion is based on the statistical studies of the number of grain agglomerates by EBSD imaging. In **Figure 5d-i** two typical agglomerates are shown (top and bottom image panels). Local orientation contrast (**Figure f,i**) and image quality (**Figure e,h**) in the middle of the agglomerate (exactly at the GB position) is stronger for the bottom panel. These parameters (image quality and local misorientation) reflect a clear concentration of the mechanical strain.[57,58] This behavior cannot be attributed to any artifact because EBSD patterns fitting quality is more than enough and corresponding low angle deviation in orientation contrast is observed. Therefore, local misorientation observed at the GBs can be a result of the mechanical strain concentration in $BiFeO_3$ films.

Summarizing previous discussion, the observed self-organization of the domain structure can be attributed to the unavoidable temperature gradients during cooling. The GBs appeared in thin film could be classified into two categories: (i) aligned boundaries which exhibit a strong structural match between the two grains, and (ii) misaligned boundaries which show slight structural mismatch with the mechanical strain localized in vicinity of GBs. Thus, grains form clear agglomerates with the cluster-like domain structure, while circumstance of the clusters represent misaligned GBs with enhanced conductivity. Nevertheless, based on the current experimental results, it is difficult to propose one single mechanism responsible for the self-organization. It is thought that the formation of mesoscale domain clusters tends to reduce electrostatic and mechanical energy of the system, while the conventional mechanisms of depolarization field and mechanical stress compensation by the formation of the domains is forbidden in superparaelectric state due to potential barrier of GBs.[42]

Assuming that the polarization reversal cannot change the position of highly conductive GBs, we suggest that the enhanced conductivity is a result of the segregation of structural defects in a similar way, how it happens at the domain walls in BFO ceramics.[25] To detangle this strain-polarization-defect segregation coupling we propose the following scenario. During the crystallization process at elevated temperatures the domain walls in a growing polar phase take their positions in places with the residual mechanical strain/stress accumulation (domain wall pinning at GBs). Further, the process of mechanical and electrical energy minimization drives defects (e.g., Bi and oxygen vacancies) to screen spontaneous polarization in the vicinity of GBs in the circumference of grain agglomerates. Both polarization charge itself and accumulated charged defects can lead to a local bending of the conduction and valence bands with corresponding enhancement of electronic conductivity. At room temperature the defects are immobile and difficult to be moved by an external electric field, but still they provide localized states for electron-hole hopping transport. This explanation is tentative, however, it reflects the main features of the observed phenomena. Nevertheless, we cannot exclude an additional contribution from non-uniformly distributed electric field arising on cooling from the crystallization temperature. This can lead to the appearance of pyroelectric charges and thus to domain self-organization.[59]

## *2.4. Finite element modelling of the properties of grain boundaries*

In order to put the proposed qualitative explanation of GBs effect on theoretical footing, we used well-known Landau-Ginsburg-Devonshire (LGD) formalism to model the interface charge and polarization distribution in polycrystalline BFO films. Schematic of the considered system consisting of a polycrystalline BFO film with a granular structure placed between the conducting AFM tip and the bottom electrode is shown in **Figure 6.** AFM tip is modelled as quasi-planar electrode separated from the film surface by a gap. "Light" and "dark" denote the crystalline regions separated by GBs which correspond to the PFM image contrast the oppositely oriented out-of-plane polarization (**Figure 3**).

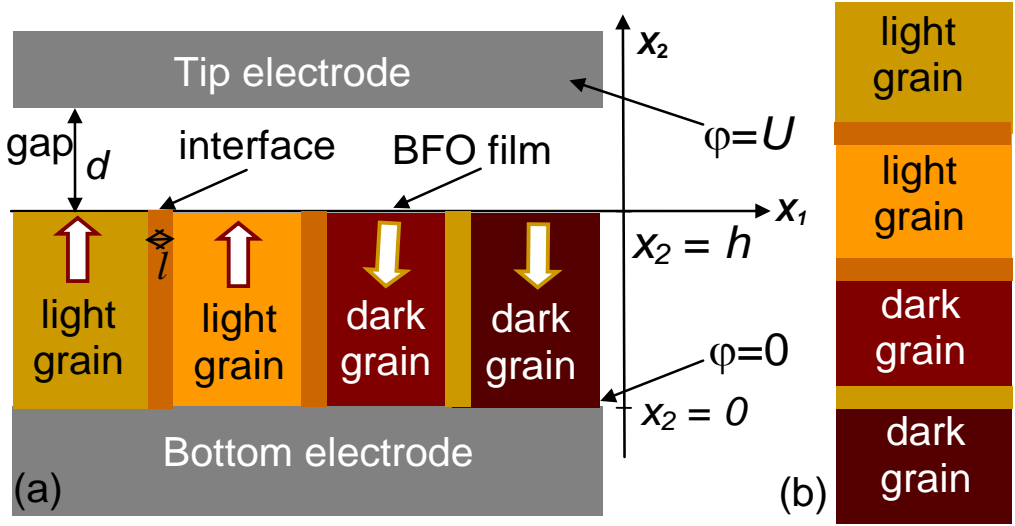

Figure 6. Schematics of the system, consisting of a polycrystalline BFO film with a granular structure (light and dark crystalline grains separated by thin GBs), placed between conducting AFM tip (modeled as quasi-planar electrode separated from the film surface by a gap) and conducting bottom electrode. (A) Cross-section view. (B) Top view. The scale is distorted, because the GBs are much thinner than the grains. White arrows schematically show the direction of the polarization normal (out-of-plane) component in the initial state.

Two vectorial order parameters, namely polarization and oxygen octahedral tilt components, $P_i$ and $\Phi_i$, are considered for the description of ferroelectric (FE) and antiferrodistortive (AFD) properties of BFO. Coupled Euler-Lagrange equations for $P_i$ and $\Phi_i$ components ($i$=1, 2, 3), obtained from the variation of LGD free energy, have the following form: [60]

$$2P_1\left(a_1 - Q_{12}(\sigma_{22} + \sigma_{33}) - Q_{11}\sigma_{11}\right) - Q_{44}(\sigma_{12}P_2 + \sigma_{13}P_3) + \left(\zeta_{11}\Phi_1^2 + \zeta_{12}(\Phi_2^2 + \Phi_3^2)\right)2P_1 + \zeta_{44}\Phi_1(\Phi_2 P_2 + \Phi_3 P_3)$$
$$+ 4a_{11}P_1^3 + 2a_{12}P_1(P_2^2 + P_3^2) + 6a_{111}P_1^5 + 2a_{112}P_1(P_2^4 + 2P_1^2 P_2^2 + P_3^4 + 2P_1^2 P_3^2) + 2a_{112}P_1 P_2^2 P_3^2$$
$$- g_{11}\frac{\partial^2 P_1}{\partial x_1^2} - g_{44}\left(\frac{\partial^2 P_1}{\partial x_2^2} + \frac{\partial^2 P_1}{\partial x_3^2}\right) - (g'_{44} + g_{12})\frac{\partial^2 P_2}{\partial x_2 \partial x_1} - (g'_{44} + g_{12})\frac{\partial^2 P_3}{\partial x_3 \partial x_1}$$
$$+ F_{11}\frac{\partial \sigma_{11}}{\partial x_1} + F_{12}\left(\frac{\partial \sigma_{22}}{\partial x_1} + \frac{\partial \sigma_{33}}{\partial x_1}\right) + F_{44}\left(\frac{\partial \sigma_{12}}{\partial x_2} + \frac{\partial \sigma_{13}}{\partial x_3}\right) = E_1$$

(1a)

$$2P_2(a_1 - Q_{12}(\sigma_{11} + \sigma_{33}) - Q_{11}\sigma_{22}) - Q_{44}(\sigma_{12}P_1 + \sigma_{23}P_3) + (\zeta_{11}\Phi_2^2 + \zeta_{12}(\Phi_1^2 + \Phi_3^2))2P_2 + \zeta_{44}\Phi_2(\Phi_1 P_1 + \Phi_3 P_3)$$
$$+ 4a_{11}P_2^3 + 2a_{12}P_2(P_1^2 + P_3^2) + 6a_{111}P_2^5 + 2a_{112}P_2(P_1^4 + 2P_2^2 P_1^2 + P_3^4 + 2P_2^2 P_3^2) + 2a_{112}P_2 P_1^2 P_3^2$$
$$- g_{11}\frac{\partial^2 P_2}{\partial x_2^2} - g_{44}\left(\frac{\partial^2 P_2}{\partial x_1^2} + \frac{\partial^2 P_2}{\partial x_3^2}\right) - (g'_{44} + g_{12})\frac{\partial^2 P_1}{\partial x_2 \partial x_1} - (g'_{44} + g_{12})\frac{\partial^2 P_3}{\partial x_3 \partial x_2}$$
$$+ F_{11}\frac{\partial \sigma_{22}}{\partial x_2} + F_{12}\left(\frac{\partial \sigma_{11}}{\partial x_2} + \frac{\partial \sigma_{33}}{\partial x_2}\right) + F_{44}\left(\frac{\partial \sigma_{12}}{\partial x_1} + \frac{\partial \sigma_{23}}{\partial x_3}\right) = E_2$$

(1b)

$$2P_3(a_1 - Q_{12}(\sigma_{11} + \sigma_{22}) - Q_{11}\sigma_{33}) - Q_{44}(\sigma_{13}P_1 + \sigma_{23}P_2) + (\zeta_{11}\Phi_3^2 + \zeta_{12}(\Phi_1^2 + \Phi_2^2))2P_3 + \zeta_{44}\Phi_3(\Phi_1 P_1 + \Phi_2 P_2)$$
$$+ 4a_{11}P_3^3 + 2a_{12}P_3(P_1^2 + P_2^2) + 6a_{111}P_3^5 + 2a_{112}P_3(P_1^4 + 2P_3^2 P_1^2 + P_2^4 + 2P_2^2 P_3^2) + 2a_{112}P_3 P_1^2 P_2^2$$
$$- g_{11}\frac{\partial^2 P_3}{\partial x_3^2} - g_{44}\left(\frac{\partial^2 P_3}{\partial x_1^2} + \frac{\partial^2 P_3}{\partial x_2^2}\right) - (g'_{44} + g_{12})\frac{\partial^2 P_1}{\partial x_3 \partial x_1} - (g'_{44} + g_{12})\frac{\partial^2 P_2}{\partial x_3 \partial x_2}$$
$$+ F_{11}\frac{\partial \sigma_{33}}{\partial x_3} + F_{12}\left(\frac{\partial \sigma_{11}}{\partial x_3} + \frac{\partial \sigma_{33}}{\partial x_3}\right) + F_{44}\left(\frac{\partial \sigma_{13}}{\partial x_1} + \frac{\partial \sigma_{23}}{\partial x_2}\right) = E_3$$

(1c)

$$(b_1 - R_{12}(\sigma_{22} + \sigma_{33}) - R_{11}\sigma_{11} + \zeta_{11}P_1^2 + \zeta_{12}(P_2^2 + P_3^2))2\Phi_1 - R_{44}(\sigma_{12}\Phi_2 + \sigma_{13}\Phi_3) + \zeta_{44}P_1(\Phi_2 P_2 + \Phi_3 P_3)$$
$$+ 4b_{11}\Phi_1^3 + 2b_{12}\Phi_1(\Phi_2^2 + \Phi_3^2) + 6b_{111}\Phi_1^5 + 2b_{112}\Phi_1(\Phi_2^4 + 2\Phi_1^2\Phi_2^2 + \Phi_3^4 + 2\Phi_1^2\Phi_3^2) + 2b_{112}\Phi_1\Phi_2^2\Phi_3^2$$
$$- \nu_{11}\frac{\partial^2 \Phi_1}{\partial x_1^2} - \nu_{44}\left(\frac{\partial^2 \Phi_1}{\partial x_2^2} + \frac{\partial^2 \Phi_1}{\partial x_3^2}\right) - (\nu'_{44} + \nu_{12})\frac{\partial^2 \Phi_2}{\partial x_1 \partial x_2} - (\nu'_{44} + \nu_{12})\frac{\partial^2 \Phi_3}{\partial x_1 \partial x_3} = 0$$

(2a)

$$(b_1 - R_{12}(\sigma_{11} + \sigma_{33}) - R_{11}\sigma_{22} + \zeta_{11}P_2^2 + \zeta_{12}(P_1^2 + P_3^2))2\Phi_2 - R_{44}(\sigma_{12}\Phi_1 + \sigma_{23}\Phi_3) + \zeta_{44}P_2(\Phi_1 P_1 + \Phi_3 P_3)$$
$$+ 4b_{11}\Phi_2^3 + 2b_{12}\Phi_2(\Phi_1^2 + \Phi_3^2) + 6b_{111}\Phi_2^5 + 2b_{112}\Phi_2(\Phi_1^4 + 2\Phi_1^2\Phi_2^2 + \Phi_3^4 + 2\Phi_2^2\Phi_3^2) + 2b_{112}\Phi_2\Phi_1^2\Phi_3^2$$
$$- \nu_{11}\frac{\partial^2 \Phi_2}{\partial x_2^2} - \nu_{44}\left(\frac{\partial^2 \Phi_2}{\partial x_1^2} + \frac{\partial^2 \Phi_2}{\partial x_3^2}\right) - (\nu'_{44} + \nu_{12})\frac{\partial^2 \Phi_1}{\partial x_1 \partial x_2} - (\nu'_{44} + \nu_{12})\frac{\partial^2 \Phi_3}{\partial x_2 \partial x_3} = 0$$

(2b)

$$(b_1 - R_{12}(\sigma_{22} + \sigma_{11}) - R_{11}\sigma_{33} + \zeta_{11}P_3^2 + \zeta_{12}(P_2^2 + P_1^2))2\Phi_3 - R_{44}(\sigma_{23}\Phi_2 + \sigma_{13}\Phi_1) + \zeta_{44}P_3(\Phi_2 P_2 + \Phi_1 P_1)$$
$$+ 4b_{11}\Phi_3^3 + 2b_{12}\Phi_3(\Phi_1^2 + \Phi_2^2) + 6b_{111}\Phi_3^5 + 2b_{112}\Phi_3(\Phi_2^4 + 2\Phi_3^2\Phi_2^2 + \Phi_1^4 + 2\Phi_1^2\Phi_3^2) + 2b_{112}\Phi_3\Phi_1^2\Phi_2^2$$
$$- \nu_{11}\frac{\partial^2 \Phi_3}{\partial x_3^2} - \nu_{44}\left(\frac{\partial^2 \Phi_3}{\partial x_2^2} + \frac{\partial^2 \Phi_3}{\partial x_1^2}\right) - (\nu'_{44} + \nu_{12})\frac{\partial^2 \Phi_1}{\partial x_1 \partial x_3} - (\nu'_{44} + \nu_{12})\frac{\partial^2 \Phi_2}{\partial x_2 \partial x_3} = 0$$

(2c)

Description and numerical values of the phenomenological coefficients $a_i$, $a_{ij}$, $a_{ijk}$, $b_i$, $b_{ij}$, $b_{ijk}$ and gradient coefficients $\nu_{ij}$ and $g_{ij}$, electrostriction $Q_{ijkl}$, rotostriction $R_{ijkl}$ and flexoelectric $F_{ijkl}$ coefficients included in equations (1)-(2), as well as LGD free energy, are summarized in Table F1 (supporting information F).

The values $\sigma_{ij}$ are elastic stresses, which satisfy the equation of state for elastic strain:

$$u_{ij} = s_{ijkl}\sigma_{kl} + Q_{ijkl}P_k P_l + R_{ijkl}\Phi_k\Phi_l + F_{ijkl}\frac{\partial P_k}{\partial x_l}\ ,$$

and mechanical equilibrium equations: $\partial\sigma_{ij}/\partial x_j = 0$. Elastic boundary conditions are the absence of mechanical displacement at the film-substrate interface $x_2 = 0$, and the absence of normal stress at the top surface, $x_2 = h$. The natural boundary conditions for polarization and tilt components to equations (1)-(2) are[60]:

$$v_{ijkl}\frac{\partial\Phi_k}{\partial x_l}n_j\bigg|_S = 0, \quad (i=1, 2, 3), \tag{3a}$$

$$\left(g_{ijkl}\frac{\partial P_k}{\partial x_l} - F_{klij}\sigma_{kl}\right)n_j = 0, \quad (i=1, 2, 3), \tag{3b}$$

where S is the film surface, $x_2 = 0$ and $x_2 = h$, and $n_j$ are the outer normal components to the surfaces. Note that following Glinka and Marton semi-microscopic model[61], we suggested that $g'_{44} + g_{12} \equiv 0$ and $v'_{44} + v_{12} \equiv 0$.

In accordance with the classical LGD theory, we assume that the coefficients $b_i$ and $a_k$ are temperature dependent in accordance with a Barrett law[60,62], and vanish at the interface $x_1 = 0$ between the grains:

$$b_i(x_1) = b_T T_{q\Phi}\left(\coth\left(\frac{T_{q\Phi}}{T}\right) - \coth\left(\frac{T_{q\Phi}}{T_\Phi}\right)\right)\tanh\left(\frac{|x_1|}{L_C}\right), \tag{4a}$$

$$a_i(x_1) = \alpha_T\left(T_{qP}\coth\left(\frac{T_{qP}}{T}\right) - T_C\right)\tanh\left(\frac{|x_1|}{L_C}\right). \tag{4b}$$

In equation (A.4a) $T_\Phi$ is the AFD transition temperature and $T_{q\Phi}$ is a characteristic temperature.[62] In equation (A.4b) $T_C$ is the Curie temperature and $T_{qP}$ is a characteristic temperature. [60,63]

Electric field components are $E_i$, which are defined by electrostatic potential in the conventional way, $E_i = -\partial\varphi/\partial x_i$. The potential satisfies Poisson equation inside the film:

$$\varepsilon_0 \varepsilon_b \frac{\partial^2 \varphi}{\partial x_i^2} = \frac{\partial P_i}{\partial x_i}, \quad 0 \le x_2 \le h, \tag{5a}$$

and Laplace equation outside the film,

$$\varepsilon_0 \varepsilon_b \frac{\partial^2 \varphi}{\partial x_i^2} = 0, \quad h \le x_2 \le \infty. \tag{5b}$$

Here $\varepsilon_b$ is the background permittivity[64], listed in Table F1 (supporting information F). Electric boundary conditions for the film on the grounded bottom electrode are zero electric potential at the electrode and potential vanishing at infinity,

$$\varphi\big|_{x_2=0} = 0, \quad \varphi\big|_{x_2\to\infty} = 0. \tag{6a}$$

The conditions of potential continuity and normal components of electric displacements equality to the screening charge $\sigma(\varphi)$ are valid at the electrically open surface of the film $x_2 = h$,

$$\varphi\big|_{x_2=h-0} = \varphi\big|_{x_2=h+0}, \quad D_2\big|_{x_2=h-0} - D_2\big|_{x_2=h+0} = \Sigma(\varphi)\big|_{x_2=h}. \tag{6b}$$

For the simplest case of (almost) complete screening of the spontaneous polarization under the absence of applied bias the electric displacement are zero in the gap, and so $D_2\big|_{x_2=h-0} = \Sigma(\varphi)\big|_{x_2=h}$.

Below we use the linear dependence of the charge density $\Sigma$ on electric potential excess $\delta\varphi = \varphi_{int}\big|_{x_2=h} - \varphi_{ext}\big|_{x_2\to\infty}$ at the surface of a ferroelectric [65,66]:

$$\Sigma[\phi] \approx -\varepsilon_0 \frac{\delta\phi}{\Lambda^*}, \qquad (6c)$$

where $\Lambda^*$ is an "effective" screening length that can vary in a wide range from e.g. 0.01 nm to 10 nm.[67] The model equation (6c) operates with "effective" screening charge, because the real space charge is distributed in the ultrathin layer near the interface[68]. The first justification of equation (6c) was proposed by Wurfel et al [69], who showed that the space charge distribution in the imperfect electrodes with nonzero screening length could be reproduced by the model in which ideal conducting electrodes are separated from the ferroelectric by a vacuum gap, and all bound and free charges are located at the interfaces. Later on Stengel et al have shown that the concept of effective screening length can be generalized for a given ferroelectric/electrode interface, at that the interfacial capacitance per unit area is proportional to $\varepsilon_0/\Lambda$ .[70,71] Introduction of the interfacial capacitance $C_{IF} = \varepsilon_0 \varepsilon_{IF} S/\Lambda$ (in a flat capacitor approximation) allows one to justify the origin of equation (6c) in a simple way, because the product $C_{IF}\varphi|_{r=R}$ is the total value of the interfacial space charge, $q = \Sigma S$, and therefore $\Sigma = \frac{C_{IF}\varphi|_{r=R}}{S} \approx -\varepsilon_0 \frac{\varphi|_{r=R}}{\Lambda}$ if the strict inequality $R \gg \Lambda$ is valid and the interfacial dielectric permittivity $\varepsilon_{IF}$ is close to unity, $\varepsilon_{IF} \approx 1$. However more likely that $\varepsilon_{IF} \gg 1$, as anticipated for an ultra-thin paraelectric passive layer, and so $\Lambda$ should be redefined as $\Lambda^* \cong \Lambda/\varepsilon_{IF}$. Equations (1)-(6) were further solved by means of finite element modeling (FEM) as a coupled problem.

The concentration of free electrons $n$ in the conductive band and the concentration of holes in a valence band $p$ of a thick polycrystalline BFO film can be estimated in the continuous levels approximation[31]:

$$n(\varphi, \sigma_{ij}) = \int_0^\infty d\varepsilon \cdot g_n(\varepsilon) f(\varepsilon - E_F + E_C(\sigma_{ij}) - e\varphi), \qquad (7a)$$

$$p(\varphi, \sigma_{ij}) = \int_0^\infty d\varepsilon \cdot g_p(\varepsilon) f(\varepsilon - E_F + E_V(\sigma_{ij}) + e\varphi), \tag{7b}$$

where $g_{n,p}(\varepsilon) \approx \sqrt{2m_{n,p}^3 \varepsilon}/(2\pi^2 \hbar^3)$ is the density of states in the effective mass approximation, $f(x) = (1 + \exp(x/k_B T))^{-1}$ is the Fermi-Dirac distribution function, $k_B = 1.3807 \times 10^{-23}$ J/K, $T$ is the absolute temperature and $e = 1.6 \times 10^{-19}$ C is the electron charge. $E_F$ is the Fermi energy level. The stress-dependent bottom of conductive band and the top of valence band

$$E_C(\sigma_{ij}) = E_{C0} + \Xi_{ij}^C \delta\sigma_{ij}, \qquad E_V(\sigma_{ij}) = E_{V0} + \Xi_{ij}^V \delta\sigma_{ij}, \tag{8}$$

where the values $E_{C0,V0}$ already include the homogeneous spontaneous stresses (if any), $\Xi_{ij}^{C,V}$ are the deformation potential tensors and $\delta\sigma_{ij} = \sigma_{ij} - \sigma_{ij}^S$. The stress tensor obeys the equation of state,

$$\sigma_{kl} = c_{ijkl} \left( u_{ij} - Q_{ijkl} P_k P_l - R_{ijkl} \Phi_k \Phi_l - F_{ijkl} \frac{\partial P_k}{\partial x_l} \right) \tag{9}$$

that contains the contributions from electrostriction ($\sim Q_{ijkl}$), rotostriction ($\sim R_{ijkl}$) and flexoelectric ($\sim F_{ijkl}$) effects. The estimations for the band gap derivative $\partial E_g / \partial \sigma_{ij} \sim 67$ meV/GPa are available for BFO.[72] The concentration of holes in the valence band can be considered in a similar way. Thus the local static conductivity $\Omega$, directly related to the experimentally measured c-AFM contrast, can be estimated as

$$\Omega = e(\mu_e n + \mu_p p), \tag{10}$$

where $\mu_e$ is the electron mobility, $\mu_p$ is the hole mobility. Hence, in accordance with the proposed model the experimentally observed conductivity enhancement at the GBs can be caused by the electric potential and elastic strain variation interfaces. It appears that the contribution of the strain variations to the conductivity modulation is dominant, because the potential changes its sign at the interface between the grains. In this simplified model we do not consider the effect of defect segregation discussed above.

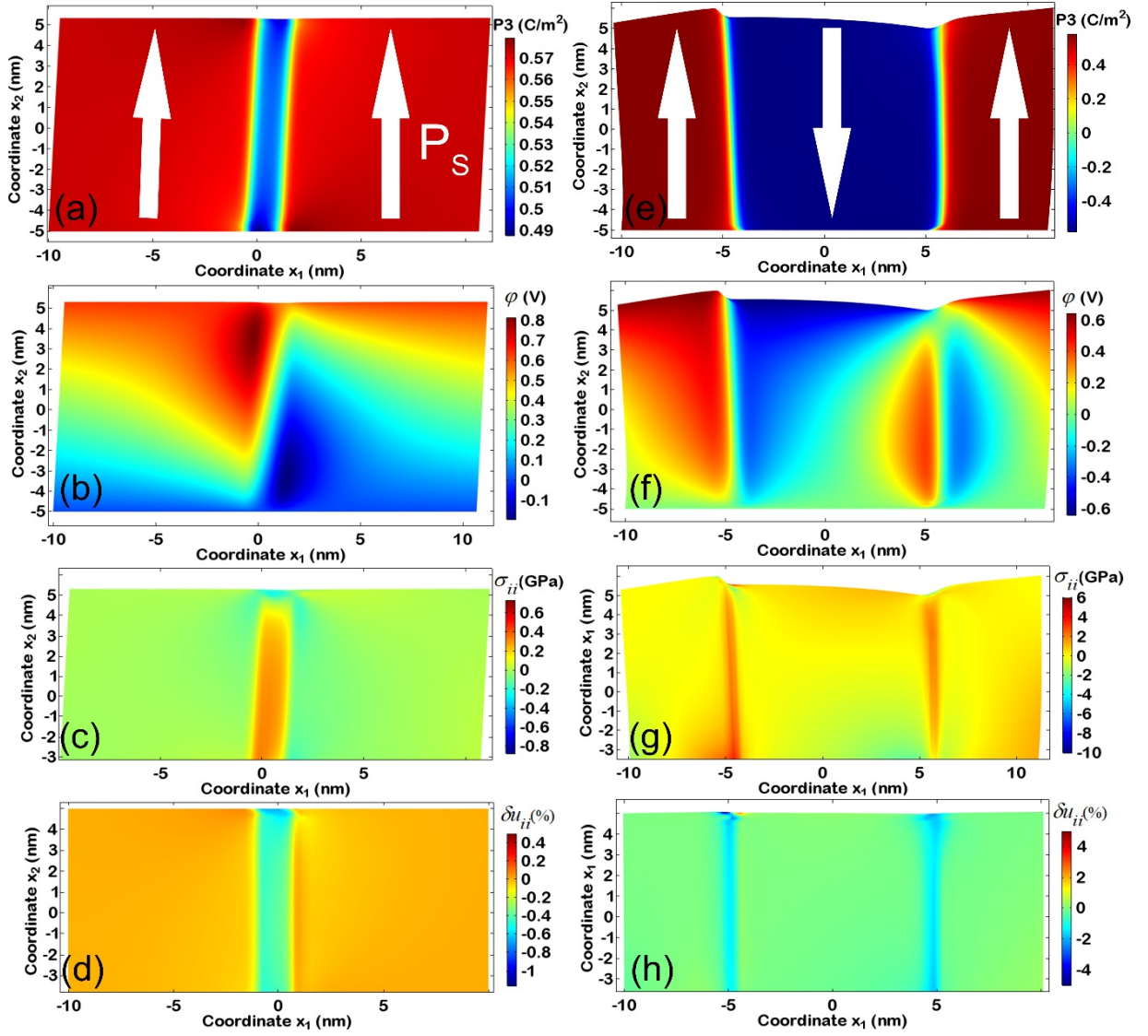

Figure 7. FEM of (a)-(d) highly conductive (polarization up-down and down-up interfaces) and (e)-(h) low conductive GBs (polarization up-down interface) in polycrystalline BFO films. (a) Distributions of out-of-plane, $P_2$, polarization components, (b) electrostatic potential, (c) hydrostatic stress near the diffuse interface $x_1 = 0$, (d) volume expansion for the polarization up-down and down-up interfaces. a) Distributions of out-of-plane, $P_2$, polarization components, (b) electrostatic potential, (c) hydrostatic stress near the diffuse interface $x_1 = 0$, (d) volume expansion for the polarization up-up interfaces.

FEM based on described model was used to calculate the polarization, tilts, electric field, surface potential and elastic stress distributions, and further analyze the c-AFM contrast across the interfaces regarding it to be proportional to local *dc* conductivity, $\Omega$. FEM results are presented in **Figure 7**; and BFO parameters used in the calculations are listed in Table F1 (Supporting Information F)**.** They have been defined earlier to fit the experimentally observed phase diagrams, domain morphology and other polar properties. [60,73,74] To prevent the domain splitting inside the grains we used small (but realistic) values of the effective screening length, $\Lambda^* \sim 0.01$ nm.

In **Figure 7e**, one can see that the distribution of polarization changes, but rather weakly in numbers, at the interface between the crystallites with almost the same direction of spontaneous polarization. The polarization profile across the interface $x_1 = 0$ is slightly inclined due to the shear strain increase near the surface, making the boundary inclined regardless the natural boundary conditions at the "nominally" straight interface $x_1 = 0$. From Figures 7f-h, corresponding changes of electric potential, elastic strains and stresses are more significant under the surface of the film, at distances about 1-5 nm. Exactly the changes lead to the 5-8 times enhancement of local static conductivity across the interface shown in **Figure 8a**.

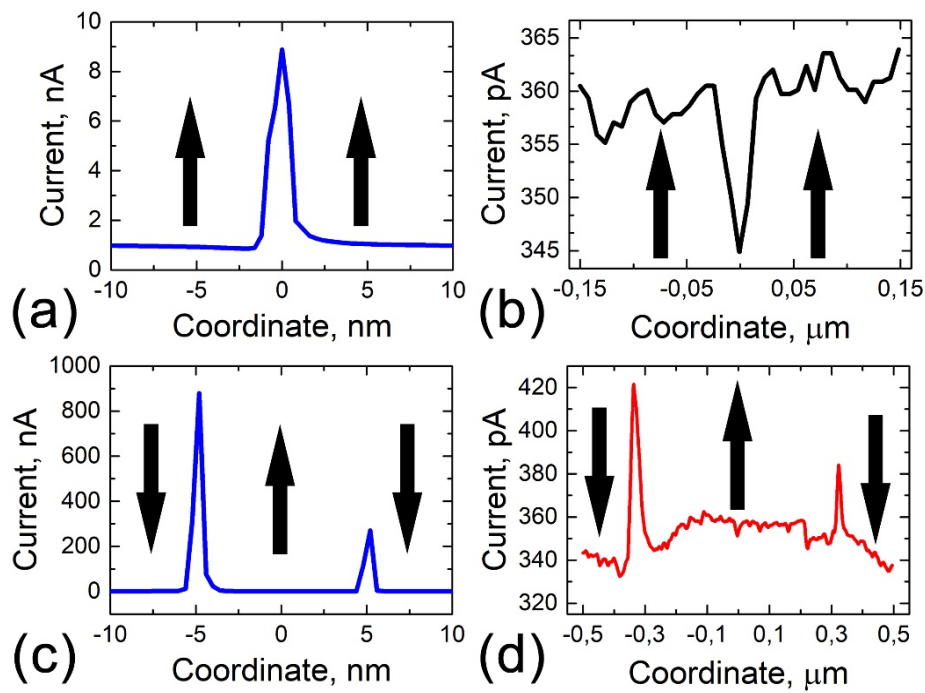

Figure 8. (a), (b) Calculated from FEM surface and (c), (d) measured by C-AFM distribution of the static conductivity variation across the (a), (c) polarization up-up and (b,d) polarization up-down and down-up interfaces.

In **Figure 7a**, one can see that the significant changes of polarization distribution at the interfaces between the crystallites with almost opposite direction of spontaneous polarization. The polarization profile across the interface $x_1 = 0$ is inclined due to the shear strain appearance at the nominally straight interface $x_1 = 0$. In **Figure 7b-d**, corresponding changes of electric potential, elastic strain and stress are significant at the surface and under the surface of the film, and the changes are mainly "bi-polar". Exactly the changes lead to the sharp enhancement (in 500 - 2000 times) of local *dc* conductivity across the interface, and, consequently to the appearance of c-AFM contrast, shown in **Figure 8b**. Note that significant strain and stress fields inevitably lead to the increase of the apparent width of the interface that corresponds to that observed experimentally through the inspection of topography (cracks appear as a result of stress concentration) and EBSD contrasts (strain localized at GBs) (**Figure 5**). Actually, we "on purpose" use a rather small "intrinsic" width $L_C$=1 nm in the calculations, to show that the apparent changes of the emerging electric and strain fields can be at least 10 times wider (~ 10 nm) across the infinitely sharp interface. Note that in the frame of considered model all asymmetries of elastic fields at up-down and down-up interfaces originate from the flexoelectric effect that breaks the symmetry in the direction perpendicular to the nominally uncharged interface and "charges" it. [31,75,76] Accumulation of the charge at the interfaces leads to the change of the Fermi level position and, correspondingly, to the enhanced conductivity at that interface. It is worth noting that in real sol-gel films the GBs can be already slightly inclined from the substrate normal and this would increase the amount of charge localized at the interfaces.

In summary, the modeling results predict that the c-AFM contrast should be different across the interfaces with different orientations of polarization vector in the adjacent grains. This is in a qualitative agreement with the obtained experimental results (**Figure 8c,d**). However, experimentally measured current at polarization up-up interfaces was either negligible, or even had an opposite direction (**Figure 8c**). This can result from the change of conductivity type and formation of p-n junction on this interface due to accumulation of different defect types such as oxygen or bismuth vacancies. This is more complicated phenomenon because of the interplay of defect chemistry, polarization and ionic/electronic transport at these naturally formed interfaces and, therefore, is out of the scope of this work. Modeling results predict the pronounced difference in the current behavior for polarization up-down and down-up interfaces (**Figure 8b**), which is in a qualitative agreement with the experimental results (**Figure 8d**).

## 3. Conclusions

In this work we demonstrated the existence of polarization-dependent conductivity along the GBs in bismuth ferrite sol-gel films. The detailed studies done by a combination of piezoresponse force and conductive atomic force microscopies revealed the existence of grain clusters with self-organized domain structure accompanied with the formation of two different type GBs: low conductive separating grains with the same polarization direction and highly conductive separating grains with antiparallel polarization. Thereby, the domain structure was organized in mesoscale clusters separated by highly conductive GBs. Based on electron backscattered diffraction, we demonstrate that many GBs are significantly strained. We hypothesized that the mechanism of such self-organization is strain-mediated and can be due to the temperature gradient during cooling after crystallization. We believe that the revealed polarization dependent conductivity at the GBs is not specific for BFO thin films, but can be a general effect in polycrystalline ferroelectrics. As such it should be taken into account for the fundamental understanding of the electrical properties of these, which are governed by the charged GBs appearing at the stage of materials synthesis.

The existence of low and highly conductive GBs may have an important impact on many macroscopic properties, e.g. dielectric permittivity and leakage current. It becomes even more important for nanosized-grain ceramics, where the influence of multiple GBs is dominant. Conductive GBs may have as well significant effect on the domain wall motion and polarization reversal in polycrystalline materials.

## 4. Experimental section

BiFeO$_3$ thin films were deposited on Pt/TiO$_2$/SiO$_2$/Si(100) substrates by spin coating using a chemical solution with 7.5 mol% of excess bismuth. The 0.16 M solution was obtained by dissolution of Bi(NO$_3$)$_3$.5H$_2$O and Fe(NO$_3$)$_3$.9H$_2$O in 2-methoxyethanol and glacial acetic acid at 50 °C for 10 min. The films were crystallized in an electric furnace (in air) at 600 ºC for 40 min to obtain a final film ~ 500 nm thick. The heating and cooling were done with 5 ºC /min rate.

The crystal structure of the BiFeO$_3$ film was characterized by X-ray diffraction (XRD) using a Rigaku Ultima IV diffractometer with CuKα (λ = 1.5406 Å) radiation. A typical perovskite structure randomly oriented without secondary phases was observed in the XRD pattern. Considering the *R3c* space group for the BiFeO$_3$ system, Rietveld refinements were conducted to analyze the XRD pattern. The lattice parameters and other structural parameters obtained for the BiFeO3 film agree with similar results for thin films in the literature. Raman measurements were performed using a confocal Raman BX51-Voyage™ with laser power of 150 mW, excitation of 785 nm and spectral resolution of 3 cm$^{-1}$. Within 13 active Raman-modes predicted by the group theory, 12 Raman modes were observed in the polycrystalline BiFeO3 film studied in the present work. Both XRD and Raman results were summarized in the supporting information material (supporting information A).

Piezoresponse Force Microscopy (PFM) was done using MFP-3D-SA (Asylum Research, Oxford Instruments, UK) scanning probe microscope with NT-MDT NSG 01 commercial tips with Pt coating, 35 nm tip radius, about 5 N/m spring constant and 150 kHz free resonance frequency. Self-developed rotational stage was used for angle-resolved PFM imaging and measuring all three spatial components of piezoelectrically driven surface displacement. The visualization of domain structure by PFM was done at 20 kHz excitation frequency that is far from first contact resonance (higher 450 kHz).

Local conductivity measurements were done in the incorporated "spreading resistance" mode of NTEGRA Aura in dark conditions and with additional illumination by focused UV diode (365 nm). The diode light was focused in about 1 cm$^2$ area with about 200 power density. Up to 10 V DC voltage was applied to NT-MDT VIT_P/Pt tips with "top-visual" geometry allowing illuminate samples, about 30 nm radius, 50 N/m spring constant and 300 kHz free resonance frequency. AFM photo-diode was screened from UV light by correspondent optical filter.

Electron backscattered diffraction (ESBD) measurements were carried out using 20 kV accelerating voltage and 5 nA electron beam current at Carl Zeiss Auriga Workstation equipped with Oxford Instruments Channel5 system. The area of 200·200 μm$^2$ were scanned with 20-50 nm step size. Electron back-scattering patterns were collected by the NordlysF+ EBSD detector and analyzed by the Flamenco Acquisition software. The local texture analysis was done by the Mambo software. The sample was covered by 2 nm carbon layer before measurements to avoid surface charging and polarization reversal under the action of electron beam.

In order to reduce the high conductivity observed in air-annealed BiFeO$_3$ films and to enable electrical characterizations, some BiFeO$_3$ films crystallized in air were subjected to an additional post annealing at 600 °C for 5 h in O$_2$ atmosphere under pressure ~ 4.0 atm. Thus, all macroscopic electrical measurements reported in this work were performed on the BFO films with lower conductivity obtained after annealing in O$_2$ atmosphere. For electrical measurements, circular gold electrodes of 0.30 mm diameter were spattered on the surface of the film using a mask. The complex dielectric permittivity ($\varepsilon^* = \varepsilon' + j\varepsilon''$) and the complex impedance ($Z^* = Z' + jZ''$) were measured at room temperature using an Agilent 4284A LCR meter in the frequency range of $10^2 – 10^6$ Hz. To discriminate the electrode polarization and GB conduction process, the dielectric data were also replotted using electric modulus formalism ($M^* = M' + jM'' = 1/\varepsilon^*$).

**Supporting Information**

Supporting Information is available from the the author.

**Acknowledgement**

This work was developed within the scope of the project CICECO-Aveiro Institute of Materials, FCT Ref. UID/CTM/50011/2019, financed by national funds through the FCT/MCTES. The equipment of the Ural Center for Shared Use "Modern nanotechnology" UrFU was used. The work was supported by Government of the Russian Federation (Act 211, Agreement 02.A03.21.0006). For the financial support, we also express our gratitude to the Brazilian agencies: Fundação de Amparo à Pesquisa do Estado de São Paulo – FAPESP (Project N° 2017/13769-1) and Conselho Nacional de Desenvolvimento Científico e Tecnológico – CNPq (Research Grant 304604/2015-1 and Project N° 400677/2014-8). This project has received funding from the European Union's Horizon 2020 research and innovation programme under the Marie Skłodowska-Curie grant agreement No 778070.

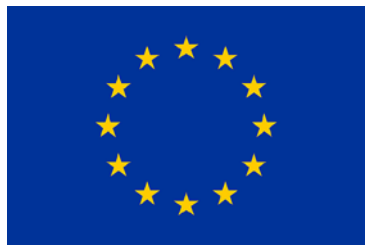

# Supporting Information

**Title** Polarization-dependent conductivity of grain boundaries in BiFeO3 thin films

*Denis Alikin[*], Yevhen Fomichov, Saulo Portes Reis, Alexander Abramov, Dmitry Chezganov, Vladimir Shur, Eugene Eliseev, Anna Morozovska[*], Eudes Araujo, and Andrei Kholkin[*]*

## A. Structural results: XRD and Raman

The crystal structure of the BiFeO$_3$ at room temperature is rhombohedrally distorted perovskite structure described within the hexagonal space group R3c. The lattice distortions ascribed to off-centering of the bismuth ions in BiFeO$_3$ leads to a spontaneous electric polarization that persists up to T$_C$ ~ 1100 K, when the rhombohedral phase shows a transition to a cubic paraelectric phase[1]. The lattice parameters obtained in the present work, $a = b = 5.5740$ Å and $c = 13.8591$ Å, are in good agreement with those parameters reported in the literature for BiFeO$_3$ single crystal[2], $a = b = 5.57874$ Å and $c = 13.8688$ Å, at room temperature.

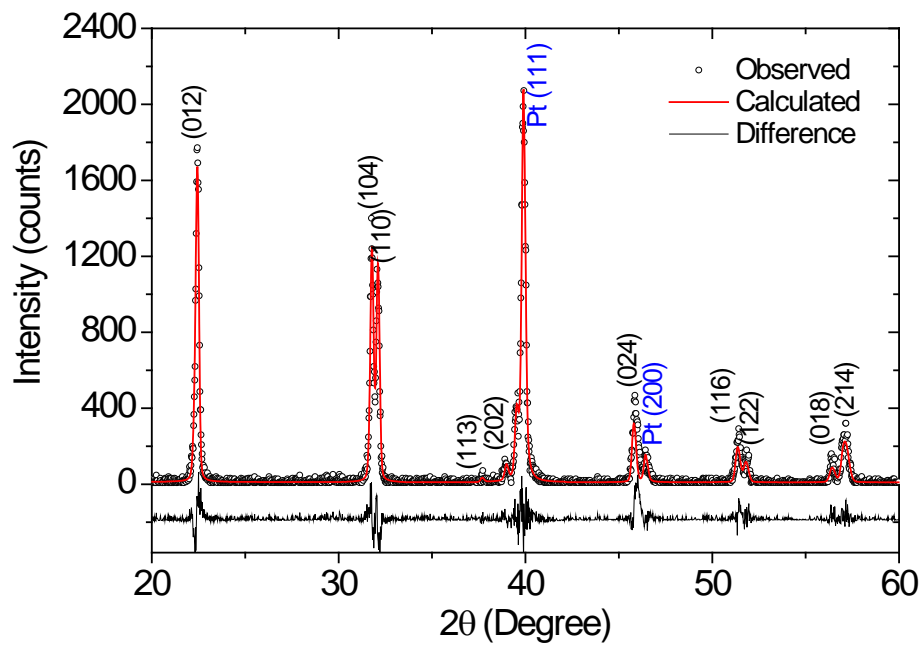

Figure A1. Observed (dots), calculated (red line) and difference (bottom line) XRD profile of the studied BiFeO$_3$ thin film.

Table A1: Refined structural parameters of BiFeO$_3$ thin film using rhombohedral space group R3c.

|  | X | Y | Z | U(Å$^2$) |
|---|---|---|---|---|
| Bi$^{3+}$ | 0.0000 | 0.0000 | 0.0000 | U$_{iso}$ = 0.078(7) |
| Fe$^{3+}$ | 0.0000 | 0.0000 | 0.2228(8) | U$_{iso}$ = 0.060(0) |
| O$^{2-}$ | 0.4366(0) | 0.0355(7) | 0.9394(4) | U$_{iso}$ = 0.022(8) |
| $a$ = 5.5740(7) Å, $b = a$, $c$ = 13.8591(2) Å and Volume = 372.91(9) Å$^3$ ||||
| $\alpha = \beta = 90°$ and $\gamma = 120°$ ||||
| $\chi^2$ = 1.6    w$_{Rp}$ = 28.4 %    R$_{exp}$ = 19.9 % ||||

The irreducible representation $\Gamma = 4A1 + 9E$ summarizes the Raman and infrared (IR) active modes[3] of the rhombohedral BiFeO$_3$ (R3c), as predicted by the group theory. Except for the E mode expected to appear below 100 cm$^{-1}$, all 12 Raman modes were observed in the polycrystalline BiFeO$_3$ film studied in the present work.

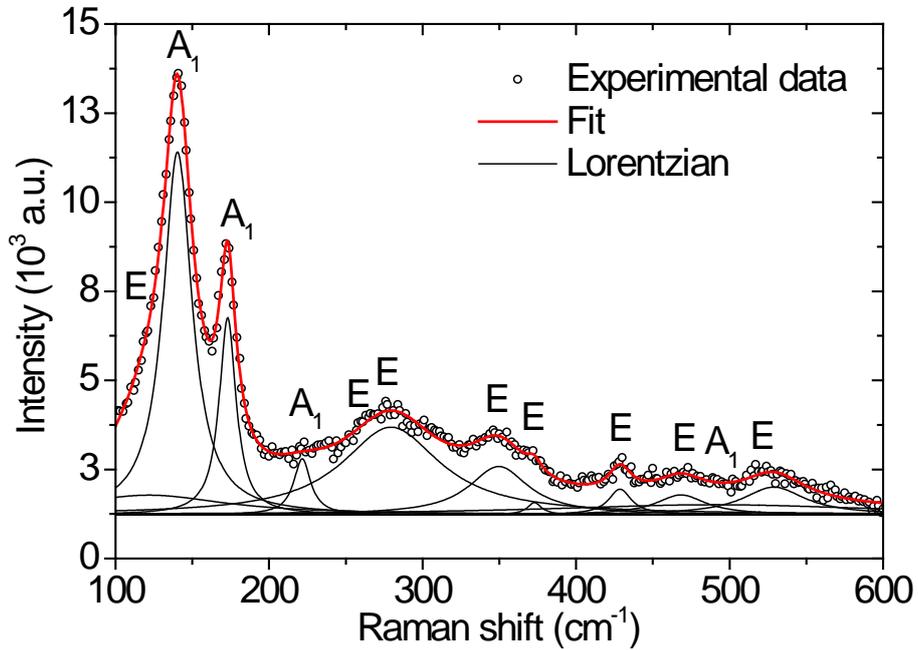

Figure A2. Room-temperature Raman spectrum of the 500-nm-thick polycrystalline rhombohedral BiFeO$_3$ thin film (open dot), full fitting (red line) and Lorentzian curves (black lines) used for fitting representing the individual Raman peaks corresponding to different phonon modes.

Table S3. Assignment and Raman modes obtained in the present work from the spectrum fit shown in Figure S2, in comparison with results from the literature for BiFeO$_3$ single crystals and epitaxial BiFeO$_3$ thin films.

| Symmetry | Present work (cm$^{-1}$) | Crystal[4] (cm$^{-1}$) | Film[3] (cm$^{-1}$) |
|---|---|---|---|
| A$_1$ | 140 | 140 | 135 |
| A$_1$ | 173 | 173 | 172 |
| A$_1$ | 219 | 220 | 218 |
| A$_1$ | 490 | | |
| E | 120 | | |
| E | 250 | 265 | 266 |
| E | 280 | 279 | 277 |
| E | 349 | 350 | 350 |
| E | 372 | 371 | 365 |
| E | 429 | | |
| E | 470 | 471 | 465 |
| E | 529 | 550 | 548 |

**B. Complex impedance and electric modulus analysis**

The complex impedance and complex electric modulus are related with the complex permittivity such as $Z^* = Z' + jZ'' = (j\omega C_0 \varepsilon^*)^{-1}$ and $M^* = M' + jM'' = (\varepsilon^*)^{-1}$. Unlike an ideal Debye-type relaxation, the non-Debye[5] relaxation in ferroelectric thin films can be described by a simple parallel RC circuit by using the Cole-Cole equation $Z^* = R[1 + (j\omega\tau)^\alpha]^{-1}$, where $\omega = 2\pi f$ is the angular frequency, $\tau$ is the relaxation time and $\alpha$ parameter ($0 < \alpha \leq 1$) indicates the deviation from the ideal Debye-type relaxation. Under these assumptions, the $Z'(\omega)$, $Z''(\omega)$, $M'(\omega)$ and $M''(\omega)$ components were evaluated for the Cole-Cole model such as:

$$Z' = \frac{R\left[1 + (\omega\tau)^\alpha \cos\left(\frac{\alpha\pi}{2}\right)\right]}{1 + 2(\omega\tau)^\alpha \cos\left(\frac{\alpha\pi}{2}\right) + (\omega\tau)^{2\alpha}} \quad \text{and} \quad Z'' = \frac{R\left[(\omega\tau)^\alpha \sin\left(\frac{\alpha\pi}{2}\right)\right]}{1 + 2(\omega\tau)^\alpha \cos\left(\frac{\alpha\pi}{2}\right) + (\omega\tau)^{2\alpha}} \quad (S1)$$

$$M' = \frac{R(\omega C_0)\left[(\omega\tau)^\alpha \sin\left(\frac{\alpha\pi}{2}\right)\right]}{1 + 2(\omega\tau)^\alpha \cos\left(\frac{\alpha\pi}{2}\right) + (\omega\tau)^{2\alpha}} \quad \text{and} \quad M'' = \frac{R(\omega C_0)\left[1 + (\omega\tau)^\alpha \cos\left(\frac{\alpha\pi}{2}\right)\right]}{1 + 2(\omega\tau)^\alpha \cos\left(\frac{\alpha\pi}{2}\right) + (\omega\tau)^{2\alpha}} \quad (S2)$$

where $R$ is the electric resistance and $C_0 = \varepsilon_0 \frac{A}{d}$ is the geometrical capacitance such as $d = 500$ nm and $A = 71$ nm². The equations (S1) and (S2) were used to fit the experimental data in the present study. The $dc$ conductivity was evaluated from $\sigma_{dc} = d/(AR)$ while the relaxation frequency from $f_{max} = 1/(2\pi\tau)$.

Table B1: Summary of parameters obtained from different fittings.

| Fit | α | $f_{max}$ (Hz) | R (Ω) | $\sigma_{dc}$ (Ω.cm)$^{-1}$ |
|---|---|---|---|---|
| Z´(ω) | 0.981 | 1185 | 748192 | $9.41 \times 10^{-8}$ |
| Z´´(ω) | 0.978 | 1175 | 753471 | $9.35 \times 10^{-8}$ |
| M´(ω) | 0.982 | 1199 | 745800 | $9.44 \times 10^{-8}$ |
| M´´(ω) | 0.983 | 1203 | 740400 | $9.51 \times 10^{-8}$ |

Equations S1 can be rewritten to describe Nyquist plot $Z''$ $versus$ $Z'$ such as:

$$\left(Z'' + \frac{R}{2} \cdot cotg\left(\frac{\alpha\pi}{2}\right)\right)^2 + \left(Z' - \frac{R}{2}\right)^2 = \frac{R^2}{4} \cdot cosec\left(\frac{\alpha\pi}{2}\right) \tag{S3}$$

Similarly, considering $M^* = j\omega C_0 Z^*$ the equations S2 can be also rewritten.

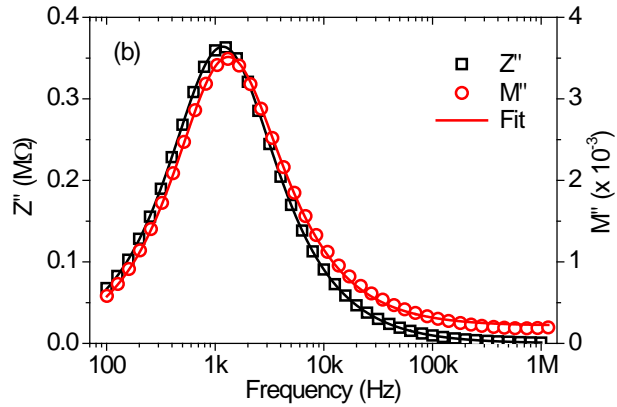

Figure B1. Frequency dependence of imaginary parts of $Z''$ and $M''$ at room temperature. The lines are theoretical fits.

## C. Local poling of BFO ceramics.

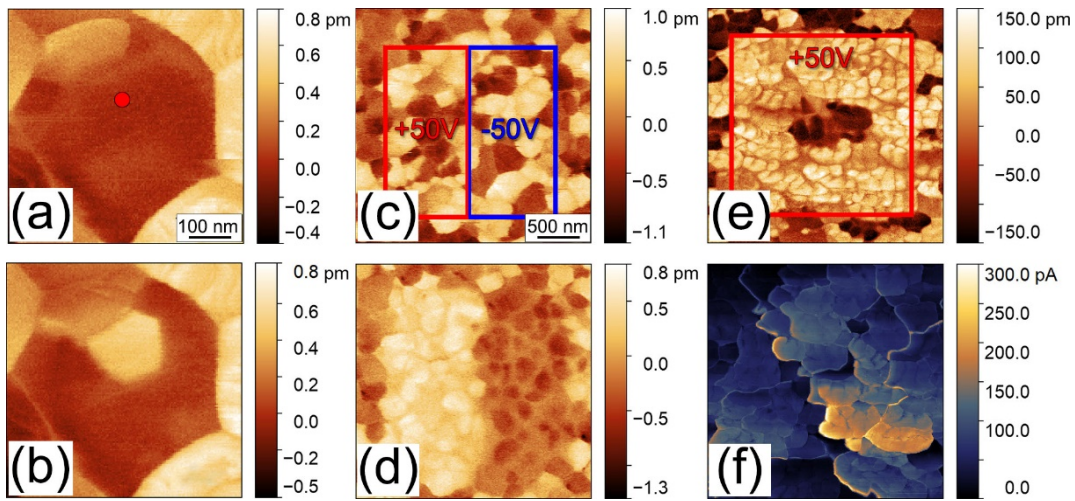

Figure C1. The local polarization switching in BFO sol-gel thin films. (a),(b) Single point switching in individual grain of the film: (a) PFM before switching (red point is a place of electric field application), (b) PFM after switching. (c)-(f) Local poling in the square area by the biased tip scanning: (c) PFM before bi-polar square poling (the poled area is denoted by corresponding color squares), (d) PFM after bi-polar square poling, (e) PFM after unipolar square poling the poled area is denoted by corresponding red color square) and (f) corresponding current image after poling. It is seen from (f) that conductivity along grain boundaries after poling doesn't correspond to the new position of the domain walls.

## D. Results of electron backscattering diffraction

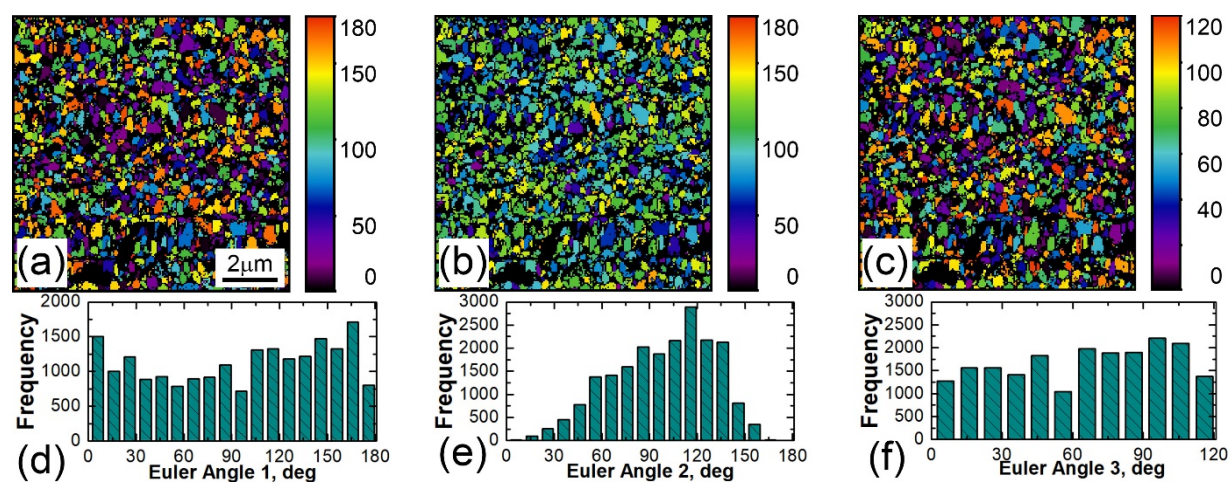

Figure D1. Orientational maps obtained by electron backscattering diffraction for the different Euler angles and corresponding histogram of orientation distribution. Self-polarization of the film is seen in out-of-plane direction.

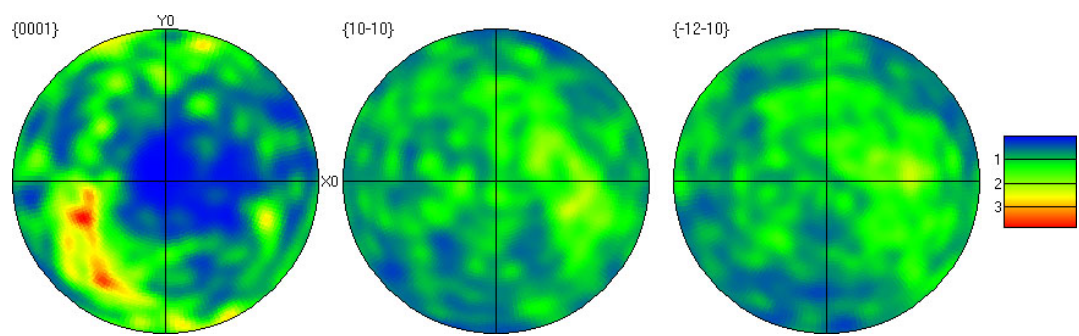

Figure D2. Back pole figures for the orientational maps at figure S1. The arrangement of the crystallites orientations in {0001} plane is non-typical for 'mechanical' texturing and thereby can be described like an electric field texturing or self-poling effect.

**E. Comparison of the current in dark condition and under UV light illumination**

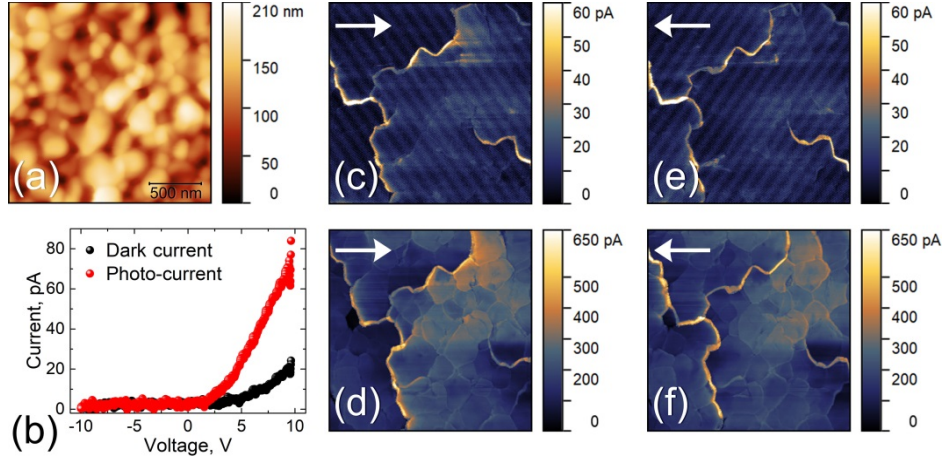

Figure E1. Comparison of normal current and photo-induced current mapping. (a) Topography, (b) current-voltage characteristics, (c) and (e) forward and backward current mapping, (d) and (f) forward and backward photo-induced current mapping. White arrow illustrate direction of the scanning.

**F. Free energy functional and parameters used in LGD calculations for BFO.**

Two vectorial order parameters, namely polarization components $P_i$ and oxygen octahedral tilts $\Phi_i$ are considered for the description of ferroelectric (FE) and antiferrodistortive (AFD) phase of BFO. For a film of thickness $h$, the bulk part of Ginzburg-Landau-Devonshire (LGD) thermodynamic potential consists of the following contributions [58]:

$$G_V = \int_S dx_1 dx_3 \int_0^h \left( \Delta G_{AFD} + \Delta G_{FE} + \Delta G_{BQC} + \Delta G_{striction} + \Delta G_{elast} + \Delta G_{flexo} \right) dx_2 \quad (F.1)$$

In equation (F.1) all contributions are consistent with the parent phase m3m symmetry in accordance with the basics of LGD approach. The AFD contribution is:

$$\Delta G_{AFD} = b_i(T)\Phi_i^2 + b_{ij}\Phi_i^2\Phi_j^2 + b_{ijk}\Phi_i^2\Phi_j^2\Phi_k^2 + v_{ijkl}\frac{\partial \Phi_i}{\partial x_k}\frac{\partial \Phi_j}{\partial x_l} \quad (F.2a)$$

In accordance with the classical Landau approach, we assume that the coefficients $b_i$ are temperature dependent. In accordance with Ref. [6][7], the dependence can be described by a Barrett law[6]: $b_i = b_T T_{q\Phi}(\coth(T_{q\Phi}/T) - \coth(T_{q\Phi}/T_\Phi))$, where $T_\Phi$ is the AFD transition temperature and $T_{q\Phi}$ is a characteristic temperature. Description and numerical values of the phenomenological coefficients $b_i$, $b_{ij}$, $b_{ijk}$ and gradient coefficients $v_{ij}$ included in equation (F.2a) can be found in Table FI, where Voight notations are used.

The FE contribution is:

$$\Delta G_{FE} = a_i(T)P_i^2 + a_{ij}P_i^2 P_j^2 + a_{ijk}P_i^2 P_j^2 P_k^2 + g_{ijkl}\frac{\partial P_i}{\partial x_k}\frac{\partial P_j}{\partial x_l} - P_i E_i \tag{F.2b}$$

In accordance with LGD approach that is well-adopted for proper and incipient ferroelectrics, the coefficients $a_k$ are temperature dependent and obeys the Barrett law, $a_k^{(P)} = \alpha_T(T_{qP}\coth(T_{qP}/T) - T_C)$, where $T_C$ is the Curie temperature and $T_{qP}$ is a characteristic temperature [58]. Description and numerical values of the phenomenological coefficients $a_i$, $a_{ij}$, $a_{ijk}$ and gradient coefficients $g_{ij}$ included in equation (F.2b) can be found in table FI.

The potential satisfies Poisson equation, $\varepsilon_0 \varepsilon_{eff}\frac{\partial^2 \varphi}{\partial x_i^2} = \frac{\partial P_i}{\partial x_i}$, where the effective dielectric permittivity, $\varepsilon_{eff} = \sum_i \varepsilon_{bi} + \varepsilon_{el}$, includes a background permittivity[8], Jahn-Teller modes and electronic contributions, which in total can be pretty high for BFO.

The biquadratic coupling energy between polarization and tilt is

$$\Delta G_{BQC} = \zeta_{ijkl}\Phi_i \Phi_j P_k P_l , \tag{F.2c}$$

As one can see, the coupling energy (F.2c) includes poorly known tensorial AFD-FE biquadratic coupling coefficients $\zeta_{44}$, $\zeta_{11}$ and $\zeta_{12}$, which have been treated as fitting parameters to experiment[9].

Electrostriction and rotostriction contributions are

$$\Delta G_{striction} = -Q_{ijkl}\sigma_{ij}P_k P_l - R_{ijkl}\sigma_{ij}\Phi_k \Phi_l \qquad (F.2d)$$

Electrostriction and rotostriction coefficients, $Q_{ijkl}$ and $R_{ijkl}$, are listed in Table FI. Elastic and flexoelectric energies are

$$\Delta G_{elast} = -s_{ijkl}\sigma_{ij}\sigma_{kl} - \frac{F_{ijkl}}{2}\left(\sigma_{ij}\frac{\partial P_k}{\partial x_l} - P_k\frac{\partial \sigma_{ij}}{\partial x_l}\right) \qquad (F.2e)$$

Here $s_{ijkl}$ are the components of elastic compliances tensor and $F_{ijkl}$ are flexoelectric tensor components (listed in Table F1).

**Table F1.** Parameters used in LGD calculations for BFO. Voight notations are used.

| Parameter | Designation | Numerical value for BFO |
|---|---|---|
| Effective permittivity | $\varepsilon_{eff} = \Sigma_i \varepsilon_{bi} + \varepsilon_{el}$ | 7-150 |
| dielectric stiffness | $\alpha_T$ ($\times 10^5$ C$^{-2}\cdot$Jm/K) | 9 |
| Curie temperature for P | $T_C$ (K) | 1300 |
| Barret temperature for P | $T_{qP}$ (K) | 800 |
| polar expansion 4$^{th}$ order | $a_{ij}$ ($\times 10^8$ C$^{-4}\cdot$m$^5$J) | $a_{11}= -13.5$, $a_{12}= 5$ |
| LGD expansion 6$^{th}$ order | $a_{ijk}$ ($\times 10^9$ C$^{-6}\cdot$m$^9$J) | $a_{111}= 11.2$, $a_{112}= -3$, $a_{123}= -6$ |
| electrostriction | $Q_{ij}$ (C$^{-2}\cdot$m$^4$) | $Q_{11}=0.054$, $Q_{12}= -0.015$, $Q_{44}=0.02$ |
| Stiffness components | $c_{ij}$ ($\times 10^{11}$ Pa) | $c_{11}=3.02$, $c_{12}= 1.62$, $c_{44}=0.68$ |
| polarization gradient coefficients | $g_{ij}$ ($\times 10^{-10}$C$^{-2}$m$^3$J) | $g_{11}=10$, $g_{12}= -7$, $g_{44}=5$ |
| AFD-FE coupling | $\xi_{ij}$ ($\times 10^{29}$ C$^{-2}\cdot$m$^{-2}$J/K) | $\xi_{11} = -0.5$, $\xi_{12} = 0.5$, $\xi_{44} = -2.6$ |
| tilt expansion 2$^{nd}$ order | $b_T$ ($\times 10^{26}\cdot$J/(m$^5$K)) | 4 |
| Curie temperature for $\Phi$ | $T_\Phi$ (K) | 1440 |
| Barret temperature for $\Phi$ | $T_{q\Phi}$ (K) | 400 |
| tilt expansion 4$^{nd}$ order | $b_{ij}$ ($\times 10^{48}$J/(m$^7$)) | $b_{11}= -24+4.5\ (\coth(300/T)-\coth(3/14))$ <br> $b_{12}= 45-4.5\ (\coth(300/T)-\coth(1/4))$ |
| tilt expansion 6$^{nd}$ order | $b_{ijk}$ ($\times 10^{70}$ J/(m$^9$)) | $b_{111}= 4.5-3.4\ (\coth(400/T)-\coth(2/7))$ <br> $b_{112}= 3.6-0.04\ (\coth(10/T)-\coth(1/130))$ |
| | | $b_{123}= 41-43.2 \left(\coth(1200/T) - \coth(12/11)\right)$ |
| tilt gradient coefficients | $v_{ij}$ ($\times 10^{11}$ J/m$^3$) | $v_{11}=2$, $v_{12}=-1$, $v_{44}=1$ |
| rotostriction | $R_{ij}$ ($\times 10^{18}$ m$^{-2}$) | $R_{11}= -1.32$, $R_{12}= -0.43$, $R_{44}=8.45$ |
| Flexoelectric coefficients | $F_{ij}$ ($\times 10^{-11}$ m$^3$/C) | $F_{11}= 2$, $F_{12}= 1$, $F_{44}= 0.5$ |
| Characteristic width of interface | $L_C$ (nm) | 1 – 5 |
| Effective screening length | $\Lambda$ (nm) | 0.01 |
| Deformation potential tensor | $\Xi_{ij}$ (meV) | Estimate for a trace $\Xi_{11}^C + \Xi_{22}^C + \Xi_{22}^C = 67$ |